\documentclass{elsarticle}
\usepackage[utf8]{inputenc}
\usepackage{graphicx}
\usepackage{amsmath}
\usepackage{caption}
\usepackage{subcaption}
\usepackage{array}
\usepackage[english]{babel}
\usepackage{natbib}

\usepackage{glossaries}

\newacronym{fem}{FEM}{finite element method}
\newacronym{fe}{FE}{finite element}
\newacronym{iga}{IGA}{isogeometric analysis}
\newacronym{cad}{CAD}{computer-aided design}
\newacronym{nurbs}{NURBS}{non-uniform rational B-splines}
\newacronym{mor}{MOR}{model order reduction}
\newacronym{tesla}{TESLA}{TeV-Energy Superconducting Linear Accelerator}
\newacronym{ttf}{TTF}{TESLA Test Facility}
\newacronym{desy}{DESY}{Deutsches Elektronen-Synchrotron}
\newacronym{fel}{FEL}{Free Electron Laser}
\newacronym{rf}{RF}{radio-frequency}
\newacronym{hom}{HOM}{higher order mode}
\newacronym{homc}{HOMC}{higher order mode coupler}
\newacronym{te}{TE}{transverse electric}
\newacronym{tm}{TM}{transverse magnetic}
\newacronym{tem}{TEM}{transverse electric and magnetic}
\newacronym{pec}{PEC}{perfect electric conducting}
\newacronym{pmc}{PMC}{perfect magnetic conducting}
\newacronym{uq}{UQ}{uncertainty quantification}
\newacronym{ebm}{EBM}{electron-beam welding}

\makeglossaries
 
\usepackage{amssymb}
\usepackage{hyperref} 

\usepackage{siunitx}
\usepackage{booktabs}

\title{Uncertainty Quantification for Maxwell's Eigenproblem based on Isogeometric Analysis and Mode Tracking}
\author{Niklas~Georg$^{1,2,3}$, Wolfgang~Ackermann$^{1}$, Jacopo~Corno$^{1,2}$ and~Sebastian~Sch\"ops$^{1,2}$\\[0.5em]
	\small$^{1}$ Technische Universit\"at Darmstadt, Institut für Theorie Elektromagnetischer Felder\\
	\small$^{2}$ Technische Universit\"at Darmstadt, Centre for Computational Engineering\\
	\small$^{3}$ Technische Universit\"at Braunschweig, Institute of Dynamics and Vibrations}
\date{\today}

\begin{document}

\begin{abstract}
The electromagnetic field distribution as well as the resonating frequency of various modes in superconducting cavities used in particle accelerators for example are sensitive to small geometry deformations. The occurring variations are motivated by measurements of an available set of resonators from which we propose to extract a small number of relevant and independent deformations by using a truncated Karhunen–Loève expansion. The random deformations are used in an expressive uncertainty quantification workflow to determine the sensitivity of the eigenmodes. For the propagation of uncertainty, a stochastic collocation method based on sparse grids is employed. It requires the repeated solution of Maxwell’s eigenvalue problem at predefined collocation points, i.e., for cavities with perturbed geometry. The main contribution of the paper is ensuring the consistency of the solution, i.e., matching the eigenpairs, among the various eigenvalue problems at the stochastic collocation points. To this end, a classical eigenvalue tracking technique is proposed that is based on homotopies between collocation points and a Newton-based eigenvalue solver. The approach can be efficiently parallelized while tracking the eigenpairs. In this paper, we propose the application of isogeometric analysis since it allows for the exact description of the geometrical domains with respect to common computer-aided design kernels, for a straightforward and convenient way of handling geometrical variations and smooth solutions. \end{abstract}


\maketitle

\section{Motivation and Introduction}

In \gls{rf} resonators, the electromagnetic fields oscillate at certain frequencies with higher magnitude than at others. The advantageous operation of such devices at selected eigenfrequencies is technically profitable used for example to set up dedicated filter components or to provide field amplitudes, which are higher than the magnitude of the excitation. In the demanding context of charged particle accelerators for example, specifically designed resonators, which are also known as cavities, can efficiently accelerate bunched charged particle beams. Essentially, the acceleration is performed in the longitudinal direction only, although specific devices can also be developed to deflect the beam in a transverse direction. The shape of the resonating structure, as well as the walls materials, determine, to a large extent, the overall performance of the entire system. As a consequence, plenty of effort has been put in the development of proper simulation tools to computationally assist scientists and engineers during the complex design process.

Because of the limitations in the available computational resources, in the early days of such simulations the main focus was exclusively put on lossless two-dimensional (2D) eigenanalysis. Meanwhile, even expensive three-dimensional (3D) models including dissipative materials or leaky boundary conditions can be handled routinely. Due to the strong dependence of the solution with respect to the geometry, a huge amount of thought and work has been spent on the precise modeling of the resonator's surface and the corresponding boundary conditions. Various discretization schemes have been applied to introduce proper computational elements ranging from a flexible irregular triangular mesh \cite{Halbach_1976aa,Rienen_1985aa} to an efficient structured rectangular variant \cite{Weiland_1984aa,Weiland_1985aa}. In both cases, special care has been put on the accurate description of the boundary surface to retain the specific convergence order of the underlying method. A further step in this direction can be found in the application of \gls{iga} \cite{Hughes_2005aa}, e.g. eigenvalue problems \cite{Hughes_2014aa}, where the geometric modeling is naturally linked to the utilized \gls{cad} kernels by using \gls{nurbs}. 
Thus, no geometry-related modeling error is introduced if the \gls{cad} model can be described by \gls{nurbs}.

\begin{figure}[t]
	\scriptsize
	\begin{subfigure}[b]{0.49\textwidth}
		\centering
		\includegraphics[]{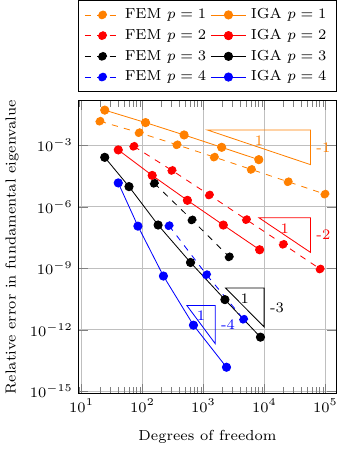}
		\caption{Different polynomial degrees $p$. For IGA, the spline regularity $C^\alpha$ is chosen as $\alpha=p-1$.}
		\label{fig:Convergence_Degree}
	\end{subfigure}\hfill
	\begin{subfigure}[b]{0.49\textwidth}
		\includegraphics[]{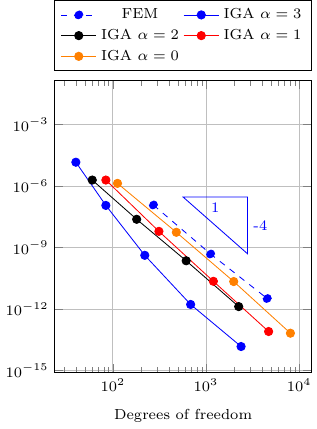}
		\caption{Fixed polynomial degree $p=4$. For IGA, the spline regularity $C^\alpha$ is varied.\\}
		\label{fig:Convergence_Regularity}
	\end{subfigure}
	\caption{Convergence of fundamental eigenvalue of Maxwell's eigenproblem on the unit square w.r.t. mesh refinement. The expected convergence orders are observed in practice. The \gls{fem} discretization is carried out using FEniCS \cite{alnaes2015} and Nédélec elements of 1st kind \cite{arnold2010finite} on a regular triangular mesh. The GeoPDEs and NURBS Octave packages are used for the \gls{iga} B-spline discretization \cite{Falco_2011aa, nurbs}.}
	\label{fig:Convergence}
\end{figure}

Another advantage of \gls{iga} is the inter-element smoothness. Quantities of interest, for example the electromagnetic fields in the vicinity of the main axis which are important for subsequent particle simulations, can be obtained with high accuracy \cite{Corno_2016aa}.  Furthermore, IGA promises faster convergence, i.e., less degrees of freedom are needed for the same accuracy, when compared to traditional polynomial approaches. As an illustration, we compare in Fig.~\ref{fig:Convergence} the accuracy of IGA discretizations of Maxwell's eigenproblem on the unit square to Nédélec elements \cite{Monk_2003aa, nedelec1980}. It can be clearly observed that higher regularity across elements leads to smaller numerical errors for smooth solutions. It shall be noted that, in both cases, there is no geometry-related modeling error. A similar comparison for a cylindrical cavity, i.e. a curved geometry, can be found in \cite{Corno_2016aa}. In this paper, it is also demonstrated that \gls{iga} is beneficial in terms of computational time. 

\begin{figure}[t]
 \begin{subfigure}[b]{0.7\textwidth}
  \centering
  \includegraphics[width=0.95\linewidth]{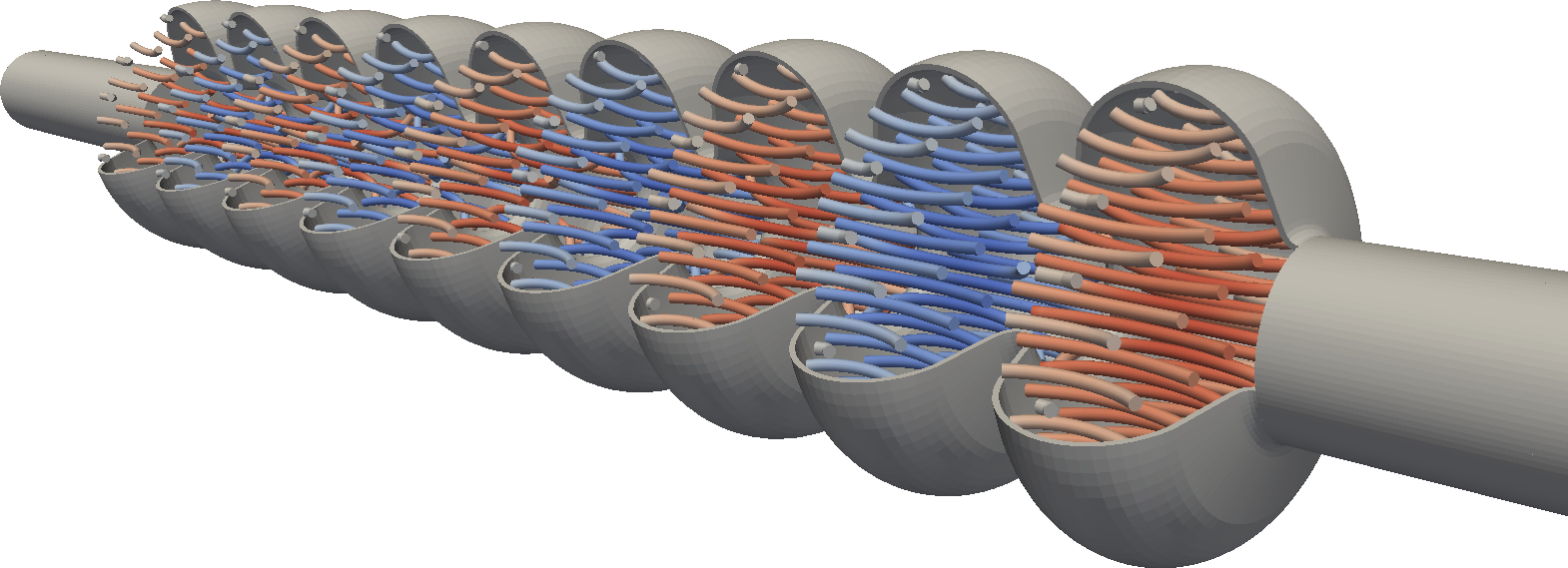}
  \caption{Electric field lines of the fundamental $\pi$-mode in the\\ \mbox{9-cell} TESLA cavity.}
  \label{fig:tesla_field_lines}
 \end{subfigure}
 \begin{subfigure}[b]{0.29\textwidth}
  \centering
  \includegraphics[width=0.95\linewidth]{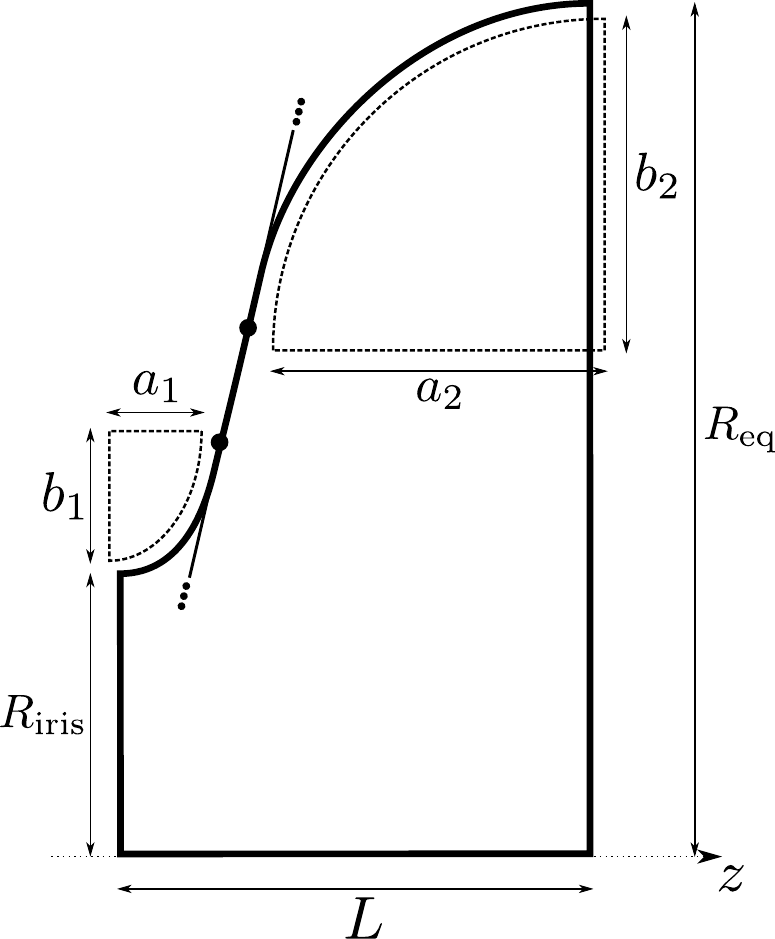}
  \caption{Specification.}
  \label{fig:images_IGA_para}
 \end{subfigure}
 \caption{Electric field solution and geometry of the TESLA cavity \cite{Aune_2000aa}.}
\end{figure}

The assembled eigenvalue formulation requires the solution of individual eigenpairs, which are also known as modes. They are characterized by a specific field distribution together with the corresponding eigenfrequency. Whereas the distribution of the electromagnetic field is connected to the determined eigenvector, the extracted eigenvalue specifies the resonating frequency and quality factor if available. In the lossy case, the modal analysis is obviously of complex nature, while in the simplified lossless approximation all calculations can be handled with real arithmetic only. One critical aspect for the manufacturing of the designed cavities is the specification of appropriate tolerances for the shape of the resonator to ensure a proper operability performance. While the sensitivity of the examined mode with respect to unavoidable geometric variations can be obtained with the help of the known sensitivity analysis, e.g. \cite{Gorgizadeh_2018aa}, a profound understanding of the corresponding distributions is obtained using the more expensive \gls{uq}, see e.g. \cite{Xiao_2007aa,Heller_2014aa}. To our best knowledge, previous works handle only tiny deformations and represent them rather naively, e.g. using non-curved elements, deforming only one layer of elements or remesh the problem without considering the mesh sensitivity. All these drawbacks are elegantly circumvented by the workflow proposed in this paper.

To study the influence of geometric boundary variations on the distribution of the electromagnetic fields and the related eigenfrequencies, the introduction of proper shape parameters has to be performed. For each cavity, which is related to a specific shape, the respective eigensolutions can be calculated independently of each other. Whereas the assignment of the simulation results with respect to the examined modes is simple for non-overlapping functions, it becomes much worse for touching or crossing graphs. In those cases, a direct evaluation of the derivatives with respect to the shape parameter is not suitable anymore and reliable eigenvalue-tracking methods have to be applied instead. In the following chapters, the necessary steps to establish a \gls{uq} workflow for electromagnetic fields in superconducting cavities using eigenvalue tracking will be formulated.

In the following section we state Maxwell's eigenvalue problem and introduce the isogeometric discretization chosen for the approximation of the fields. In section~\ref{sec:uq} we present the uncertainty quantification methods used for the simulation and their relation to the geometry parametrization. Given the necessity to properly categorize the eigensolutions for each uncertain configuration, we present the eigenvalue tracking method in section~\ref{sec:tracking}. Finally, we show the application of the proposed scheme to the simple case of the pillbox cavity and to a real world example, i.e. the \gls{tesla} cavity (see Fig.~\ref{fig:tesla_field_lines} and~\ref{fig:images_IGA_para}). In the last section we give some concluding remarks.  
\section{Numerical Approximation of Maxwell's Eigenproblem}

\subsection{Maxwell's Eigenvalue Problem}

Given a bounded and simply connected domain $\Omega \in \ensuremath{\rm I\!R}^3$, with Lipschitz continuous boundary $\partial\Omega$,
the electromagnetic fields in the source-free, 
time-harmonic case, are governed by Maxwell's equations:
\begin{equation}\label{eq:Maxwell-time-harmonic}
\begin{aligned}
\ensuremath{\nabla \times \ensuremath{\mathbf{E}}}       &= -j\omega\ensuremath{\mu_0}\ensuremath{\mathbf{H}} && \text{in }\Omega\\
\ensuremath{\nabla \times \ensuremath{\mathbf{H}}}       &= j\omega\ensuremath{\varepsilon_0}\ensuremath{\mathbf{E}} && \text{in }\Omega\\
\ensuremath{\nabla \cdot \ensuremath{\varepsilon_0}\ensuremath{\mathbf{E}}} &= 0                     && \text{in }\Omega\\
\ensuremath{\nabla \cdot \ensuremath{\mu_0}\ensuremath{\mathbf{H}}}  &= 0                     && \text{in }\Omega,
\end{aligned}
\end{equation}
where $\ensuremath{\mathbf{E}}$ and $\ensuremath{\mathbf{H}}$ denote the electric and magnetic field strength, respectively, and we assume the electric permittivity $\ensuremath{\varepsilon_0}$
and magnetic permeability $\ensuremath{\mu_0}$ of vacuum. We further apply
\gls{pec} boundary conditions
\begin{equation}\label{eq:pecBC}
\begin{aligned}
\ensuremath{\mathbf{E}}\times\ensuremath{\mathbf{n}} &= 0                     && \text{on }\partial\Omega\\
\ensuremath{\mathbf{H}}\cdot\ensuremath{\mathbf{n}}  &= 0                     && \text{on }\partial\Omega.
\end{aligned}
\end{equation}

Starting from Maxwell's equations, one can derive the classical 
eigenvalue formulation by eliminating \ensuremath{\mathbf{H}} 
from~\eqref{eq:Maxwell-time-harmonic}: 
find the wave number $k:=\omega \sqrt{\mu_0\epsilon_0} \in \ensuremath{\rm I\!R}$ and $\ensuremath{\mathbf{E}} \neq 0$ such that
\begin{equation}\label{eq:Maxwell-eig-cont}
\begin{aligned}
\ensuremath{\nabla \times \left(\ensuremath{\nabla \times \ensuremath{\mathbf{E}}}\right)} &= k^2 \ensuremath{\mathbf{E}} && \text{in }\Omega\\
\ensuremath{\nabla \cdot \ensuremath{\mathbf{E}}} &= 0 && \text{in }\Omega\\
\ensuremath{\mathbf{E}}\times\ensuremath{\mathbf{n}} &= 0 && \text{on }\partial\Omega.
\end{aligned}
\end{equation}

The standard variational formulation of~\eqref{eq:Maxwell-eig-cont} is: 
find $k\in \ensuremath{\rm I\!R}$ with $\ensuremath{\mathbf{E}} \in \ensuremath{\mathbf{H}_0\left(\textbf{curl};\Omega\right)}$ such that
\begin{equation}\label{eq:Maxwell-eig-var}
 \left(\ensuremath{\nabla \times \ensuremath{\mathbf{E}}},\ensuremath{\nabla \times \ensuremath{\mathbf{v}}}\right) 
  = k^2 \left(\ensuremath{\mathbf{E}},\ensuremath{\mathbf{v}}\right) \quad \forall \ensuremath{\mathbf{v}} \in \ensuremath{\mathbf{H}_0\left(\textbf{curl};\Omega\right)}, 
\end{equation}
where we introduced the function space $\ensuremath{\mathbf{H}_0\left(\textbf{curl};\Omega\right)}$ of square-integrable vector fields with square-integrable \textbf{curl}. For further information on function spaces in the
context of Maxwell's equations, the reader is referred to \cite{Monk_2003aa}.
The $0$ subscript in $\ensuremath{\mathbf{H}_0\left(\textbf{curl};\Omega\right)}$ indicates vanishing tangential 
components on the boundary $\partial\Omega$ since we are imposing 
\gls{pec} boundary conditions.

In order to be able to numerically solve~\eqref{eq:Maxwell-eig-var} we 
deduce its discrete counterpart: following a Galerkin procedure, 
we introduce a sequence of finite-dimensional spaces $V_h \subset \ensuremath{\mathbf{H}_0\left(\textbf{curl};\Omega\right)}$ 
and we restrict~\eqref{eq:Maxwell-eig-var} to it: 
find $\ensuremath{k_h} \in \ensuremath{\rm I\!R}$ with $\ensuremath{\mathbf{E}}_h \in V_h$ such that
\begin{equation}\label{eq:Maxwell-eig-var-discrete}
\left(\ensuremath{\nabla \times \ensuremath{\mathbf{E}}_h},\ensuremath{\nabla \times \ensuremath{\mathbf{v}}}\right) 
  = \ensuremath{k_h}^2 \left(\ensuremath{\mathbf{E}}_h,\ensuremath{\mathbf{v}}\right) \quad \forall \ensuremath{\mathbf{v}} \in V_h.
\end{equation}

Whatever the choice of the functional space $V_h$, 
$\ensuremath{\mathbf{E}}_h$ can be expressed in terms of a set of basis functions 
$\left\lbrace \ensuremath{\mathbf{v}_{j}} \right\rbrace_{j=1}^{\ensuremath{N_{\text{dof}}}}$:
\begin{equation}
\ensuremath{\mathbf{E}}_h = \sum_{j=1}^{\ensuremath{N_{\text{dof}}}} e_j \ensuremath{\mathbf{v}_{j}}.
\end{equation}
The discrete solution of Maxwell's eigenvalue problem is then 
obtained by solving the generalized eigenvalue problem
\begin{equation}\label{eq:galerkin-substitution}
\ensuremath{\mathbf{K}} \ensuremath{\mathbf{e}} = \ensuremath{k_h}^2 \ensuremath{\mathbf{M}} \ensuremath{\mathbf{e}},
\end{equation}
where $\ensuremath{\mathbf{K}}$ and $\ensuremath{\mathbf{M}}$ are the stiffness and mass 
matrix for vacuum respectively, given by
\begin{equation}\label{eq:matrix-def}
\ensuremath{\mathbf{K}}_{ij} = \left(\ensuremath{\nabla \times \ensuremath{\mathbf{v}_{j}}}, \ensuremath{\nabla \times \ensuremath{\mathbf{v}_{i}}}\right), \quad
\ensuremath{\mathbf{M}}_{ij}  = \left(\ensuremath{\mathbf{v}_{j}}, \ensuremath{\mathbf{v}_{i}}\right).
\end{equation}

There exist multiple choices for the space $V_h$ and its 
basis although care must be taken to ensure a proper approximation 
of $\ensuremath{\mathbf{H}\left(\textbf{curl};\Omega\right)}$. For the \gls{fem} case we refer the interested reader
to \cite{Boffi_2010aa}. In the following section we will 
follow~\cite{Buffa_2010aa} and introduce a curl-conforming  \gls{iga}
discretization.

\subsection{Isogeometric Analysis}

\gls{iga} is a technique that was introduced by Hughes et 
al.~\cite{Hughes_2005aa} with the aim of allowing the exact 
representation of geometries defined via \gls{cad} software,
independently of the level of refinement of the computational grid.
It is clear that, since the geometry directly affects the resonant
properties of \gls{rf} cavities, the exact description
of the simulation domain is strongly desired, even more so in the case of 
shape sensitivity analysis and \gls{uq}.

In order to achieve this goal, the same classes of basis functions
that are commonly used for geometry description in \gls{cad} 
software, are used for the representation of solution vector fields. 
Such basis functions are the so called B-splines and \gls{nurbs}. For the definition and construction of such basis, we refer the interested reader to~\cite{de-Boor_2001aa,Piegl_1997aa}.

The use of \gls{iga} in cavity simulation has been shown to be
beneficial both in terms of accuracy and of overall reduction 
of the computational cost~\cite{Corno_2016aa, Corno_2015ac}.
For an overview of the applicability of \gls{iga} to other
electromagnetic problems, we refer to \cite{Bontinck_2017ag}.

\begin{figure}[t]
  \centering
  \includegraphics[]{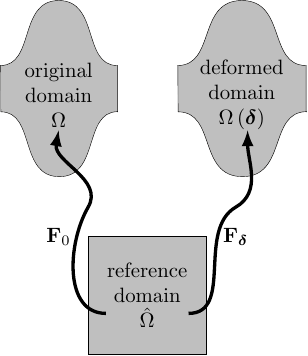}
  \caption{Mapping from reference to original and deformed physicals domain in dependence of the
  parameter $\ensuremath{\boldsymbol\delta}$.}
  \label{fig:images_IGA_mapping}
\end{figure}

We assume that the domain $\Omega$ can be exactly described 
through a smooth mapping $\ensuremath{\mathbf{F}}_0$ obtained from the
\gls{cad} software, see Fig.~\ref{fig:images_IGA_mapping}:
\begin{equation}
\ensuremath{\mathbf{F}}_0 : \ensuremath{\hat{\Omega}} \rightarrow \Omega,
\end{equation}
where $\ensuremath{\hat{\Omega}}$ is a parametric domain (e.g. the unit cube).
We furthermore assume that $\ensuremath{\mathbf{F}}_0$ is piecewise smoothly
invertible.

We define the discretization space $V_h$ in a general way
\begin{equation}
\begin{aligned}
V_h &:= \left\lbrace \ensuremath{\mathbf{v}} \in \ensuremath{\mathbf{H}\left(\textbf{curl};\Omega\right)} :\ensuremath{\hat{\ensuremath{\mathbf{v}}}}=\iota\left(\ensuremath{\mathbf{v}}\right) \in \ensuremath{\hat{V}}_h\right\rbrace\\
    &\equiv \left\lbrace \ensuremath{\mathbf{v}} \in \ensuremath{\mathbf{H}\left(\textbf{curl};\Omega\right)} :\ensuremath{\mathbf{v}}=\iota^{-1}\left(\ensuremath{\hat{\ensuremath{\mathbf{v}}}}\right), \ensuremath{\hat{\ensuremath{\mathbf{v}}}} \in \ensuremath{\hat{V}}_h\right\rbrace,
\end{aligned}
\end{equation}
where $\ensuremath{\hat{V}}_h$ is a discrete space on the parametric domain
and $\iota$ is a so-called pull-back function defined from 
the parametrization $\ensuremath{\mathbf{F}}_0$. Let 
$\lbrace\hat{\ensuremath{\mathbf{v}_{}}}_j\rbrace_{j=1}^{\ensuremath{N_{\text{dof}}}}$ be a basis for
$\ensuremath{\hat{V}}_h$, the set
$\lbrace\ensuremath{\mathbf{v}_{j}}\rbrace_{j=1}^{\ensuremath{N_{\text{dof}}}} \equiv \lbrace\iota^{-1}(\hat{\ensuremath{\mathbf{v}_{}}}_j)\rbrace_{j=1}^{\ensuremath{N_{\text{dof}}}}$
is a basis for $V_h$ that can be substituted
in~\eqref{eq:galerkin-substitution}-\eqref{eq:matrix-def}
to obtain a finite-dimensional generalized eigenvalue problem.

We indicate as 
$S^{p_1,p_2,p_3}_{\alpha_1,\alpha_2,\alpha_3}(\ensuremath{\hat{\Omega}})$
the standard space of \mbox{B-Splines} of degree $p_i$ along dimension
$i$, with regularity $C^{\alpha_i}$, constructed on the 
reference domain $\ensuremath{\hat{\Omega}}$. Following~\cite{Buffa_2010aa} 
the discrete space $\ensuremath{\hat{V}}_h$ is chosen as
\begin{equation}
\ensuremath{\hat{V}}_h := S^{p_1-1,p_2,p_3}_{\alpha_1-1,\alpha_2,\alpha_3}
  \times S^{p_1,p_2-1,p_3}_{\alpha_1,\alpha_2-1,\alpha_3}
  \times S^{p_1,p_2,p_3-1}_{\alpha_1,\alpha_2,\alpha_3-1}.
\end{equation}
It can be proven that basis functions in this space 
obey to a discrete de Rham sequence, thus being suitable candidates
for the discretization of $\ensuremath{\mathbf{H}\left(\textbf{curl};\hat{\Omega}\right)}$.

The discrete space $V_h$ in the physical domain is obtained
through a curl-conforming `Piola' transformation~\cite{Monk_2003aa}:
\begin{equation}
V_h := \left\lbrace \ensuremath{\mathbf{v}} : \left(\mathrm d\ensuremath{\mathbf{F}}_0\right)^{\!\top}\left(\ensuremath{\mathbf{v}}\circ\ensuremath{\mathbf{F}}_0^{-1}\right)\in\ensuremath{\hat{V}}_h\right\rbrace,
\end{equation}
where we indicate with $\mathrm d\ensuremath{\mathbf{F}}_0$ the Jacobian of the parametrization.

\subsection{Domain Parametrization}

Given a B-Spline (or \gls{nurbs}) basis $\lbrace B_i \rbrace$ of 
$S^{p_1,p_2,p_3}_{\alpha_1,\alpha_2,\alpha_3}(\ensuremath{\hat{\Omega}})$,
the mapping $\ensuremath{\mathbf{F}}$ is constructed using a set of control points 
$\ensuremath{\mathbf{P}}_{i}$ in the physical space as:
\begin{equation}\label{eq:iga-map}
\ensuremath{\mathbf{F}} = \sum_{i} \ensuremath{\mathbf{P}}_i B_i.
\end{equation}
The points $\ensuremath{\mathbf{P}}_i$ represent the so-called control mesh.

It is straightforward to show~\cite{Piegl_1997aa} that, 
given definition~\eqref{eq:iga-map}, \gls{cad} objects are 
invariant under affine transformations such as translations, 
scalings and rotations, i.e. to apply the transformation to the 
object it is sufficient to apply it to the control mesh.
Furthermore, $\ensuremath{\mathbf{F}}$ is smooth with respect to the movement 
of the control points.

Let us now consider the domain $\Omega$ and its parametrization
$\ensuremath{\mathbf{F}}$ to be given in terms of a $N_{\ensuremath{\boldsymbol\delta}}$-vector of parameters
$\ensuremath{\boldsymbol\delta}$, i.e. $\Omega$ becomes $\ensuremath{\Omega\left(\ensuremath{\boldsymbol\delta}\right)}$ and the mapping
\begin{equation}\label{eq:iga-map-uq}
\ensuremath{\mathbf{F}_{\ensuremath{\boldsymbol\delta}}}:\ensuremath{\hat{\Omega}} \rightarrow \ensuremath{\Omega\left(\ensuremath{\boldsymbol\delta}\right)}.
\end{equation}
The shape variability can be reinterpreted as the variability
of the control mesh:
\begin{equation}
\ensuremath{\mathbf{F}_{\ensuremath{\boldsymbol\delta}}} = \sum_i \ensuremath{\mathbf{P}}_i(\ensuremath{\boldsymbol\delta}) B_i,
\end{equation}
and by following the steps described in the previous section
in the case of a parametrized domain, one derives
the parametrized generalized eigenvalue problem:
\begin{equation}\label{eq:galerkin-substitution-p}
\ensuremath{\mathbf{K}}(\ensuremath{\boldsymbol\delta}) \ensuremath{\mathbf{e}}(\ensuremath{\boldsymbol\delta}) = k_h^2(\ensuremath{\boldsymbol\delta}) \ensuremath{\mathbf{M}}(\ensuremath{\boldsymbol\delta}) \ensuremath{\mathbf{e}}(\ensuremath{\boldsymbol\delta})
\end{equation}
where $\ensuremath{k_h}(\ensuremath{\boldsymbol\delta})$ denotes the numerical approximation of the eigenvalue $k(\ensuremath{\boldsymbol\delta})$ for each parameter $\ensuremath{\boldsymbol\delta}$. 
Suppressing for brevity the $\ensuremath{\boldsymbol\delta}$ dependency, we denote the discrete eigenpairs with $(\ensuremath{k_{h,j}},\ensuremath{\ensuremath{\ensuremath{\mathbf{e}}}_{j}})$ where $j=1,\ldots,N_\text{J}$. 
In the following, we assume the parametric dependency of the discrete eigenpairs $(\ensuremath{k_{h,j}}, \ensuremath{\ensuremath{\ensuremath{\mathbf{e}}}_{j}})$ and their continuous counterparts $(k_j, \ensuremath{\mathbf{E}}_j)$ on $\ensuremath{\boldsymbol\delta}$ to be sufficiently smooth. 
\section{Uncertainty Quantification}\label{sec:uq}
\begin{figure*}[ht]
	\centering
	\begin{subfigure}[c]{0.32\textwidth}
		\centering
		\includegraphics[]{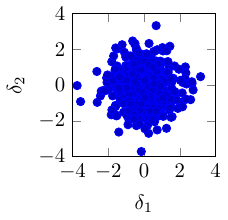}
		\caption{Monte Carlo sampling.}		
	\end{subfigure}
	\begin{subfigure}[c]{0.32\textwidth}
		\centering
		\includegraphics[]{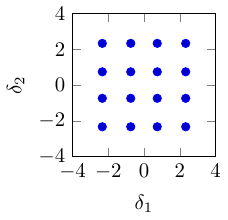}
		\caption{Tensor grid.}
	\end{subfigure}
	\begin{subfigure}[c]{0.32\textwidth}
		\centering
		\includegraphics[]{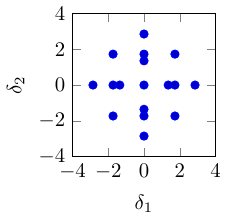}
		\caption{Sparse grid.}
	\end{subfigure}
	\caption{Choice of collocation points for the stochastic collocation method for a 2-dimensional parameter space.}
	\label{fig:grids}
\end{figure*}
In operation, the geometry of a particular TESLA cavity will differ from the design shown in Fig.~\ref{fig:images_IGA_para}: the manufacturing process, tuning and multiphysical effects as cooling, Lorentz detuning and microphoning will introduce geometry variations \cite{Xiao_2007aa, Deryckere_2012aa,Schmidt_2014aa,Corno_2016aa}. In the simplest case, these variations can be expressed in terms of the design parameters, see Fig.~\ref{fig:images_IGA_para}, i.e., 
\begin{equation}
	\ensuremath{\boldsymbol\delta}_\text{design} = \left[a_1, a_2, b_1, b_2, L, R_\textrm{iris}, R_\textrm{eq} \right]^{\!\top}
\end{equation}
and lead to an explicit definition of the parametrized geometry mapping $\ensuremath{\mathbf{F}}_{\ensuremath{\boldsymbol\delta}_\text{design}}$. However, in more realistic scenarios the variation of a particular geometry is unknown.

Let us assume that we have measured $M$ times the geometric variations and that each observation can be described by an $N$-dimensional vector \begin{equation}\mathbf{x}^{(m)} = \left[x^{(m)}_1, x^{(m)}_2, ..., x^{(m)}_N\right]^{\!\top}, \quad m=1,\ldots,M.\end{equation}
And let us further assume that all components $x^{(m)}_n$ are realizations of normally distributed random variables 
\begin{equation}
	X_n\sim\mathcal{N}(\mu_n,\sigma_n^2),\quad n = 1,\dots,N,
\end{equation}
possibly mutually correlated. Expectation and variance are denoted by $\mu_n$ and $\sigma_n^2$ respectively and can be obtained from the $M$ measurement samples.

We propose the discrete Karhunen–Loève expansion (or equivalently principal component analysis \cite{Xiu_2010aa}) to convert the observations into a set of linearly uncorrelated variables. \gls{uq} methods can then be applied to investigate the sensitivity of the eigenmodes to those variables. 

\subsection{Karhunen–Loève}
\label{sec:kl}
Let us collect all the observations in the matrix 
\begin{equation}\label{eq:obs_mat_T}
\mathbf{T}=\left[\mathbf{x}^{(1)}, \mathbf{x}^{(2)}, \hdots, \mathbf{x}^{(M)} \right]^{\!\top},
\end{equation}
whose columns are the random variables and whose rows are the measured observations. We will denote by $\mathbf{C}\in{\ensuremath{\rm I\!R}}^{N \times N}$ its covariance matrix
\begin{equation}\mathbf{C} := \frac{1}{M-1} \sum_{m=1}^{M} \left(\ensuremath{\mathbf{x}}^{(m)}-\boldsymbol{\mu}\right)\left(\ensuremath{\mathbf{x}}^{(m)}-\boldsymbol{\mu}\right)^{\!\top},\end{equation}
where $\boldsymbol{\mu}$ is the sample mean
\begin{equation}\boldsymbol{\mu}=\frac{1}{M}\sum_{m=1}^{M} \ensuremath{\mathbf{x}}^{(m)}.\end{equation}
We search for an eigendecomposition of the covariance matrix of the type
\begin{equation}
\mathbf{C}=\mathbf{V}\boldsymbol{\Sigma}\mathbf{V}^{\!\top},
\end{equation}
where the columns of $\mathbf{V}$ are the orthonormal right eigenvectors of $\mathbf{C}$, and $\boldsymbol{\Sigma}=\mathrm{diag}(\sigma_1,\ldots,\sigma_{N})$ is a diagonal matrix whose elements are the corresponding eigenvalues. Without loss of generality, we will further assume the columns of $\mathbf{V}$ to be ordered such that $\sigma_1\ge\ldots\ge\sigma_{N}$.

Now, only the $N_\text{t}\ll N$ most significant contributions are extracted by approximating
\begin{equation}\mathbf{C}\approx \mathbf{C}_\text{t}:=\mathbf{V}_\text{t} \boldsymbol{\Sigma}_\text{t} \mathbf{V}_\text{t}^{\!\top},\end{equation}
where $\boldsymbol{\Sigma}_\text{t}=\mathrm{diag}(\sigma_1,\ldots,\sigma_{N_\text{t}})$ and $\mathbf{V}_\text{t}$ is a $(N\times N_\text{t}$)-matrix with the corresponding eigenvectors.

Finally, the $N$ variations $X_n$ can be approximated by a low-dimensional representation \cite{DElia_2013aa}
\begin{align}		
	\begin{bmatrix}
		X_1 \\ \vdots\\X_{N}
	\end{bmatrix} \approx
	\begin{bmatrix}
		\mu_{1}\\\vdots\\\mu_{N}
	\end{bmatrix} 
	+ \mathbf{V}_\text{t} \boldsymbol{\Sigma}^{1/2}_\text{t}
	\begin{bmatrix}
		\ensuremath{\delta}_1\\\vdots\\\ensuremath{\delta}_{N_\text{t}}
	\end{bmatrix}.
\end{align}
that is based on normally distributed parameters $\ensuremath{\boldsymbol\delta}\sim\mathcal{N}(\ensuremath{\mathbf{0}},\pmb{\mathbb{I}})$. With respect to the original vector of parameters $\ensuremath{\boldsymbol\delta}_\text{design}$, $\ensuremath{\boldsymbol\delta}$ may not have a physical meaning, but has a significantly smaller dimension. Reducing the size of the parameter space has a great impact in the efficiency of a stochastic collocation method.

Let us now consider the case of a \gls{nurbs} domain specified by the mapping~\eqref{eq:iga-map}. Given $\ensuremath{\boldsymbol\delta}$, to describe the deformed geometry~\eqref{eq:iga-map-uq} we will firstly construct the  $N_\text{t}+1$ B-Spline interpolating curve mappings~\cite{de-Boor_2001aa}
\begin{align}
\ensuremath{\mathbf{F}}^{\mu} &= \sum_{i} \ensuremath{\mathbf{P}}^{\mu}_i B_i\\
\ensuremath{\mathbf{F}}^{j}   &= \sum_{i} \ensuremath{\mathbf{P}}^{j}_i B_i \quad \forall j=1,\dots,N_\text{t},
\end{align}
where $\ensuremath{\mathbf{F}}^{\mu}$ interpolates the mean deformation vector $[\mu_1, \cdots, \mu_{N}]$ and $\ensuremath{\mathbf{F}}^{j}$ is an interpolating curve through the points given by the $j$-th column of the matrix $\mathbf{V}_\text{t} \boldsymbol{\Sigma}^{1/2}_\text{t}$. Unique interpolations can be obtained by requiring additional properties as a certain polynomial degree and minimal curvature~\cite{Piegl_1997aa}. Both $\ensuremath{\mathbf{F}}^{\mu}$ and the $\ensuremath{\mathbf{F}}^{j}$ share the same basis functions as the geometry mapping $\ensuremath{\mathbf{F}}_{\ensuremath{\boldsymbol\delta}_\text{design}}$ and differ only in their control mesh. The control points defining the deformed geometry are then given by
\begin{equation}
\ensuremath{\mathbf{P}}_i(\ensuremath{\boldsymbol\delta}) = \ensuremath{\mathbf{P}}_i + \ensuremath{\mathbf{P}}^{\mu}_i + \sum_{j=1}^{N_\text{t}} \ensuremath{\delta}_j \ensuremath{\mathbf{P}}^{j}_i.
\end{equation}

\subsection{Stochastic Collocation}

Quantities of interest are statistical measures about the eigenmodes with respect to the influence of the vector $\ensuremath{\boldsymbol\delta}$ of random variables. In particular, we aim for expectation and variance of the $j$-th eigenfrequency $f_j=k_j/(2\pi\sqrt{\ensuremath{\varepsilon_0}\ensuremath{\mu_0}})$,
\begin{subequations}\label{eq:statistical_measures}
	\begin{align} 
	\mathbb{E}[f_j] &= \int_{-\infty}^{\infty} f_j(\ensuremath{\boldsymbol\delta})  \rho(\ensuremath{\boldsymbol\delta}) \, \mathrm d\ensuremath{\boldsymbol\delta}\\
	\operatorname{var}[f_j] &= \int_{-\infty}^{\infty} (f_j(\ensuremath{\boldsymbol\delta})-\mathbb{E}[f_j])^2 \, \rho(\ensuremath{\boldsymbol\delta}) \, \mathrm d\ensuremath{\boldsymbol\delta},
	\end{align}
\end{subequations}
where $\rho(\ensuremath{\boldsymbol\delta})$ denotes the joint probability density of $\ensuremath{\boldsymbol\delta}$.

In order to approximate \eqref{eq:statistical_measures} numerically, a stochastic collocation method \cite{Babuska_2010aa} is applied which can be summarized as follows: based on the assumption that the parametric dependency $f_j(\ensuremath{\boldsymbol\delta})$ is smooth,  
we construct a polynomial approximation based on Lagrangian interpolation
\begin{align} \label{eq:lagrange}
f_j(\ensuremath{\boldsymbol\delta})\approx \sum_{k=1}^{N_\text{k}} f_{h,j}(\ensuremath{\boldsymbol\delta}_k) l_k(\ensuremath{\boldsymbol\delta}), && l_m(\ensuremath{\boldsymbol\delta}_n) = \begin{cases} 1 & \text{if } m=n\\ 0 & \text{if } m\neq n\end{cases}
\end{align}
where $l_k$ are multivariate Lagrange polynomials constructed by tensorization of univariate Lagrange polynomials and $\{\ensuremath{\boldsymbol\delta}_k\}_{k=1}^{N_\text{k}}$ is the set of corresponding collocation points (tensor grid). The eigenfrequencies 
\begin{align}
f_{h,j}(\ensuremath{\boldsymbol\delta}_k)=\frac{\ensuremath{k_{h,j}}(\ensuremath{\boldsymbol\delta}_k)}{2\pi\sqrt{\ensuremath{\varepsilon_0}\ensuremath{\mu_0}}}.
\end{align} are obtained by solving \eqref{eq:galerkin-substitution-p} for each set of discrete parameters $\ensuremath{\boldsymbol\delta}_k$. This requires rerunning the simulation $N_\text{k}$ times for slightly different geometries. No modifications of the code are necessary and hence the method can be considered as non-intrusive. Finally, inserting \eqref{eq:lagrange} into \eqref{eq:statistical_measures} yields, for the $j$-th eigenfrequency:
\begin{subequations}
	\label{eq:quadrature}
	\begin{align}
	\mathbb{E}[f_j] &\approx \sum_{k=1}^{N_\text{k}} f_{h,j}(\ensuremath{\boldsymbol\delta}_k) w_k\\
	\operatorname{var}[f_j] & \approx \sum_{k=1}^{N_\text{k}} (f_{h,j}(\ensuremath{\boldsymbol\delta}_k)-\mathbb{E}[f_j])^2w_k,
	\end{align}
\end{subequations}
 with weights 
\begin{equation}
w_k=\int_{-\infty}^{\infty} l_k(\ensuremath{\boldsymbol\delta}) \rho(\ensuremath{\boldsymbol\delta}) \, \mathrm d\ensuremath{\boldsymbol\delta}.
\end{equation}
The integrals can be evaluated exactly using Gauss-Hermite	quadrature for each independent parameter $\ensuremath{\delta}_1, \dots, \ensuremath{\delta}_{N_\text{t}}$. The quadrature nodes are also used as collocation points $\ensuremath{\boldsymbol\delta}_k$.

The stochastic collocation method \eqref{eq:quadrature} based on tensor grids is known to be very efficient for a small number of random variables but suffers strongly from the curse of dimensionality \cite{Nobile_2008aa}. The number of collocation points $N_\text{k}$ grows exponentially with the dimension of the parametric space $N_\text{t}$. This problem can be mitigated by the sparse grid approach \cite{Bungartz_2004aa},\cite{Smolyak_1963aa}. Linear combinations of tensor grids with different degrees for different parameters $\ensuremath{\delta}_1, \dots \ensuremath{\delta}_{N_\text{t}}$ lead to more efficient interpolation and cubature rules for a (moderately) large stochastic space \cite{Novak_1999aa}. An illustration of stochastic collocation grids is given in Fig.~\ref{fig:grids}.

In each of the collocation points, eigenvalue problem \eqref{eq:galerkin-substitution} is solved. The natural ordering of the eigenpairs according to their eigenvalue, however, might not be sufficient to ensure consistency since even small changes in the domain can cause eigenvalue crossing or separation/degeneration of modes. To be able to correctly place the frequencies $f_j$ for each parametric point $\ensuremath{\boldsymbol\delta}_k$, an eigenvalue tracking algorithm is applied.
  
\section{Eigenvalue Tracking}\label{sec:tracking}
The following procedure is adapted from \cite{Lui_1997aa}.
To be able to track the eigenvalues, we construct an algebraic homotopy between the quadrature points in the parameter space in such a way that:
\begin{subequations}\label{eq:homotopy}
\begin{align}
	\ensuremath{\mathbf{K}}_{k,\ensuremath{t}} &:=(1-t)\ensuremath{\mathbf{K}}({\ensuremath{\boldsymbol\delta}_0})+\ensuremath{t}\ensuremath{\mathbf{K}}({\ensuremath{\boldsymbol\delta}_k})\\
	\ensuremath{\mathbf{M}}_{k,\ensuremath{t}}  &:=(1-t)\ensuremath{\mathbf{M}} ({\ensuremath{\boldsymbol\delta}_0})+\ensuremath{t}\ensuremath{\mathbf{M}} ({\ensuremath{\boldsymbol\delta}_k})
\end{align}
\end{subequations}
where $t\in[0,1]$ and $\ensuremath{\boldsymbol\delta}_0$ is a given start configuration in the parameter space (e.g. corresponding to an unperturbed geometry). We further assume that at $t=0$ the modes are not degenerated and that there is no bifurcation from $t=0$ to $t=1$. Other homotopies could be analogously defined, e.g. such that the eigenvalue problems for $0\le\ensuremath{t}\le1$ correspond to actual geometries. However, the algebraically motivated choice \eqref{eq:homotopy} allows the straightforward definition of the derivatives, 
\begin{subequations}\label{eq:mat_derivatives}
\begin{align}
\ensuremath{\mathbf{K}}_{k,\ensuremath{t}}'&:=\frac{\partial}{\partial\ensuremath{t}}\ensuremath{\mathbf{K}}_{k,\ensuremath{t}}=\ensuremath{\mathbf{K}}({\ensuremath{\boldsymbol\delta}_k})-\ensuremath{\mathbf{K}}({\ensuremath{\boldsymbol\delta}_0}) \\
\ensuremath{\mathbf{M}}_{k,\ensuremath{t}}'&:=\frac{\partial}{\partial\ensuremath{t}}\ensuremath{\mathbf{M}}_{k,\ensuremath{t}}=\ensuremath{\mathbf{M}}({\ensuremath{\boldsymbol\delta}_k})-\ensuremath{\mathbf{M}}({\ensuremath{\boldsymbol\delta}_0})
\end{align}
\end{subequations}
Now, all the quantities depending on the vector $\ensuremath{\boldsymbol\delta}$, can then be viewed as depending on the scalar quantity $\ensuremath{t}$ which makes the following one-dimensional analysis feasible.
For each collocation point $k$ and eigenpair $j$, the eigenvalue problem \eqref{eq:galerkin-substitution} can be rewritten as a linear system of equations with a normalization constraint using a suitable vector $\ensuremath{\mathbf{c}}$
\begin{equation}
\begin{bmatrix}
\ensuremath{\mathbf{K}}_{\ensuremath{t}} \ensuremath{\ensuremath{\ensuremath{\mathbf{e}}}_{\ensuremath{t}}} - \ensuremath{\lambda_{\ensuremath{t}}} \ensuremath{\mathbf{M}}_{\ensuremath{t}} \ensuremath{\ensuremath{\ensuremath{\mathbf{e}}}_{\ensuremath{t}}} \\
\ensuremath{\mathbf{c}}^{\!\top} \ensuremath{\ensuremath{\ensuremath{\mathbf{e}}}_{\ensuremath{t}}} - 1
\end{bmatrix} = \begin{bmatrix}
0 \\	 0
\end{bmatrix}, \label{eq:eigprob}
\end{equation}
where $\ensuremath{\lambda_{\ensuremath{t}}}:=\bigl(2\pi f_{h,j}\bigl(\ensuremath{\boldsymbol\delta}_{k}(t)\bigr) \sqrt{\ensuremath{\varepsilon_0} \ensuremath{\mu_0}}\bigr)^2$ and $\ensuremath{\ensuremath{\ensuremath{\mathbf{e}}}_{\ensuremath{t}}} := \ensuremath{\ensuremath{\ensuremath{\mathbf{e}}}_{j}}\bigl(\ensuremath{\boldsymbol\delta}_k(\ensuremath{t})\bigr)$. 
The subscripts $j$ and $k$ addressing eigenpair number and collocation point are suppressed for easier readability.

\begin{figure}
\centering
\includegraphics[]{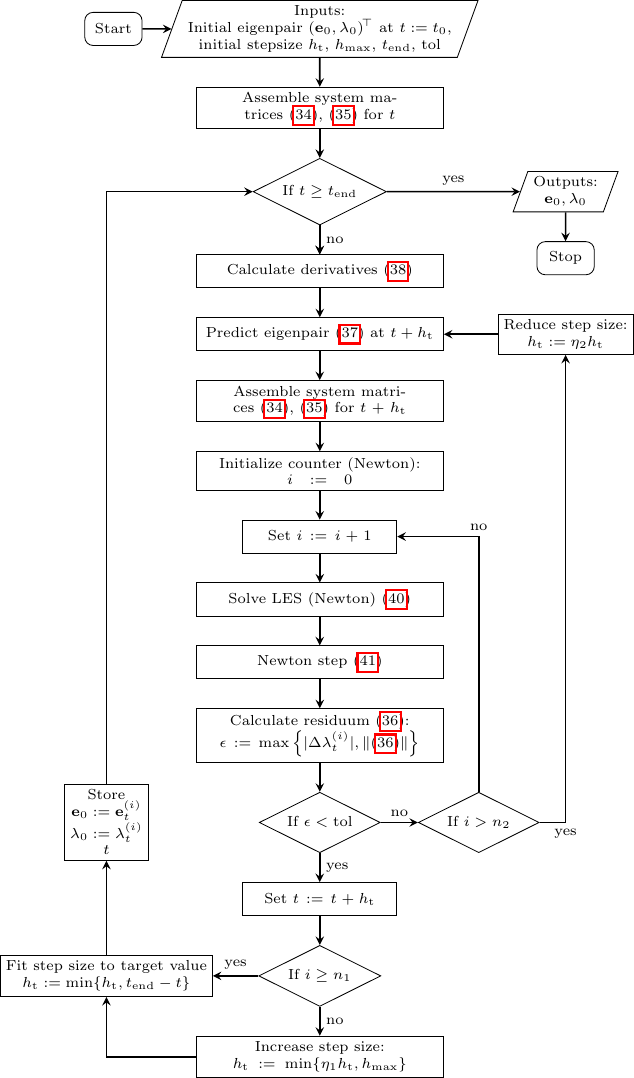} 
\caption{Flowchart describing the proposed eigenvalue tracking method. The parameters of the stepsize control are chosen as $n_1=3$, $\eta_1=1.1$, $n_2=5$, $\eta_2=\frac 2 3$.}
\label{fig:Flowchart_Tracking}
\end{figure}

Let us consider a Taylor expansion at $\ensuremath{t}_0$ along $\ensuremath{t}$
\begin{align}
	\begin{bmatrix}
	\ensuremath{\tilde{\ensuremath{\ensuremath{\mathbf{e}}}}_{\ensuremath{t}}}\\
	\ensuremath{\tilde{\lambda}_{\ensuremath{t}}}
	\end{bmatrix}=\begin{bmatrix}
	\ensuremath{\ensuremath{\mathbf{e}}}_0 \\
	\lambda_0
	\end{bmatrix}+(\ensuremath{t}-\ensuremath{t}_0)\begin{bmatrix}
	\ensuremath{\ensuremath{\mathbf{e}}}'_0 \\
	\lambda'_0 \label{eq:prediction}
	\end{bmatrix} 
\end{align}
to obtain a first-order approximation for the eigenpair $[\tilde{\ensuremath{\ensuremath{\mathbf{e}}}}_{\ensuremath{t}}^{\!\top}, \tilde{\lambda}_{\ensuremath{t}}]^{\!\top} \approx [\ensuremath{\ensuremath{\ensuremath{\mathbf{e}}}_{\ensuremath{t}}}^{\!\top}, \ensuremath{\lambda_{\ensuremath{t}}} ]^{\!\top}$ at $\ensuremath{\boldsymbol\delta}_k(\ensuremath{t})$. The derivatives $[{\ensuremath{\ensuremath{\mathbf{e}}}_0'}^{\!\top},\lambda_0']^{\!\top}$ of the eigenvalues and eigenvectors are obtained from \eqref{eq:eigprob} which is differentiated with respect to the parameter $\ensuremath{t}$. The resulting linear system 
\begin{align}
	\begin{bmatrix}
	\ensuremath{\mathbf{K}}_0 - \lambda_0\ensuremath{\mathbf{M}}_0  &\! -\ensuremath{\mathbf{M}}_0\ensuremath{\ensuremath{\mathbf{e}}}_0 \\
	\ensuremath{\mathbf{c}}^{\!\top} &\! 0
	\end{bmatrix}\begin{bmatrix}
	\ensuremath{\ensuremath{\mathbf{e}}}_0'\\\lambda_0'
	\end{bmatrix}=\begin{bmatrix}
	-\ensuremath{\mathbf{K}}_0'\ensuremath{\ensuremath{\mathbf{e}}}_0+\lambda_0\ensuremath{\mathbf{M}}_0'\ensuremath{\ensuremath{\mathbf{e}}}_0 \\0
	\end{bmatrix} \label{eq:eigprobderiv}
\end{align}
is then solved with respect to the eigenvalue and eigenvector derivatives. 
The predicted eigenpair $[\tilde{\mathbf{e}}^{\!\top}_{\ensuremath{t}}, \tilde \lambda_{\ensuremath{t}}]^{\!\top}$ can be used as an initial guess 
\begin{align}
   \begin{bmatrix}
   \ensuremath{\ensuremath{\ensuremath{\mathbf{e}}}_{\ensuremath{t}}}^{(0)}\\
   \ensuremath{\lambda_{\ensuremath{t}}}^{(0)}
   \end{bmatrix} :=
   \begin{bmatrix}
   \tilde{\ensuremath{\ensuremath{\mathbf{e}}}}_{\ensuremath{t}}\\
   \tilde \lambda_{\ensuremath{t}}
   \end{bmatrix}
\end{align}
for the Newton-Raphson method in order to obtain the exact eigenpair $[\ensuremath{\ensuremath{\ensuremath{\mathbf{e}}}_{\ensuremath{t}}}^{\!\top}, \ensuremath{\lambda_{\ensuremath{t}}}]^{\!\top}$. 
In the $i$-th iteration of the Newton-Raphson method, the linear system of equations to be solved is
\begin{align}
	\begin{bmatrix}
	\ensuremath{\mathbf{K}}_\ensuremath{t}-\ensuremath{\lambda_{\ensuremath{t}}}^{(i-1)}\ensuremath{\mathbf{M}}_\ensuremath{t} && -\ensuremath{\mathbf{M}}_\ensuremath{t}\ensuremath{\ensuremath{\ensuremath{\mathbf{e}}}_{\ensuremath{t}}}^{(i-1)} \\
	\ensuremath{\mathbf{c}}^{\!\top} && 0
	\end{bmatrix}
	\begin{bmatrix}
	\Delta\ensuremath{\ensuremath{\ensuremath{\mathbf{e}}}_{\ensuremath{t}}}^{(i)}\\
	\Delta\ensuremath{\lambda_{\ensuremath{t}}}^{(i)}
	\end{bmatrix} =\nonumber\\-\begin{bmatrix}
	\ensuremath{\mathbf{K}}_\ensuremath{t}\ensuremath{\ensuremath{\ensuremath{\mathbf{e}}}_{\ensuremath{t}}}^{(i-1)} -\ensuremath{\lambda_{\ensuremath{t}}}^{(i-1)}\ensuremath{\mathbf{M}}_\ensuremath{t}\ensuremath{\ensuremath{\ensuremath{\mathbf{e}}}_{\ensuremath{t}}}^{(i-1)}\\
	\ensuremath{\mathbf{c}}^{\!\top} \ensuremath{\ensuremath{\ensuremath{\mathbf{e}}}_{\ensuremath{t}}}^{(i-1)}-1
	\end{bmatrix} \label{eq:NewtonLGS}.
\end{align} 
The increments $\Delta \ensuremath{\ensuremath{\ensuremath{\mathbf{e}}}_{\ensuremath{t}}}^{(i)}$ and $\Delta\ensuremath{\lambda_{\ensuremath{t}}}^{(i)}$ improve the prediction by assigning
\begin{align}
	\begin{bmatrix}
	\ensuremath{\ensuremath{\ensuremath{\mathbf{e}}}_{\ensuremath{t}}}^{(i)}\\
	\ensuremath{\lambda_{\ensuremath{t}}}^{(i)}
	\end{bmatrix} := 
	\begin{bmatrix}
	\ensuremath{\ensuremath{\ensuremath{\mathbf{e}}}_{\ensuremath{t}}}^{(i-1)}\\
	\ensuremath{\lambda_{\ensuremath{t}}}^{(i-1)}	
	\end{bmatrix} + 
	\begin{bmatrix}
	\Delta\ensuremath{\ensuremath{\ensuremath{\mathbf{e}}}_{\ensuremath{t}}}^{(i)}\\
	\Delta\ensuremath{\lambda_{\ensuremath{t}}}^{(i)}	
	\end{bmatrix}.\label{eq:newtonstep}
\end{align}
As a termination criterion for the Newton-Raphson method, the norm of the residual of eigenvalue problem \eqref{eq:eigprob} can be considered as well as the norm of $\Delta \ensuremath{\lambda_{\ensuremath{t}}}^{(i)}$. 
The required number of Newton iterations can then be used for a stepsize control in order to increase the efficiency and robustness of the algorithm. The proposed tracking algorithm is illustrated by the flowchart in Fig.~\ref{fig:Flowchart_Tracking}.

It shall be noted that the assumption of non-degeneracy of the eigenmodes at $t=0$ is made for simplicity only. The algorithm could be generalized in order to address the case of an intersection of eigenvalues at $t=0$ by using Ojalvo's method \cite{ojalvo, dailey} for computing the derivatives instead of \eqref{eq:eigprobderiv}. 
The numerical efficiency of the presented tracking method could also be further improved, e.g.  by using a simplified Newton method \cite{Deuflhard_2004aa}, by employing a higher order Taylor approximation in the prediction step \eqref{eq:prediction} or by tracking a subspace of eigenpairs and matching the eigenpairs based on a correlation coefficient \cite{Jorkowski_2018aa, Yang_2007ab}.   
\section{Application}\label{sec:application}

In this section, we present the application of the eigenvalue tracking method to cavities with uncertain geometries. As a clear indication of the necessity of the tracking, we first analyze the test case of a cylindrical cavity with uncertain radius. Stochastic collocation is then applied to study the sensitivity of a \gls{tesla} type cavity \cite{Aune_2000aa} to eccentric deformation of its 9 cells. In both cases the discretization is carried out with \gls{iga} using GeoPDEs and NURBS Octave packages \cite{Falco_2011aa, nurbs}.

\subsection{Pillbox Cavity}
\begin{figure}
	\centering
	\begin{subfigure}[c]{0.495\textwidth}
		\centering
		\includegraphics[width=0.87\linewidth]{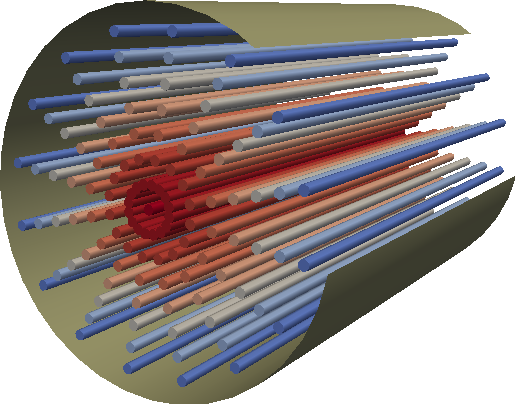}
		\caption{Electric field lines.}
	\end{subfigure}
	\begin{subfigure}[c]{0.495\textwidth}
		\centering
		\includegraphics[width=0.87\linewidth]{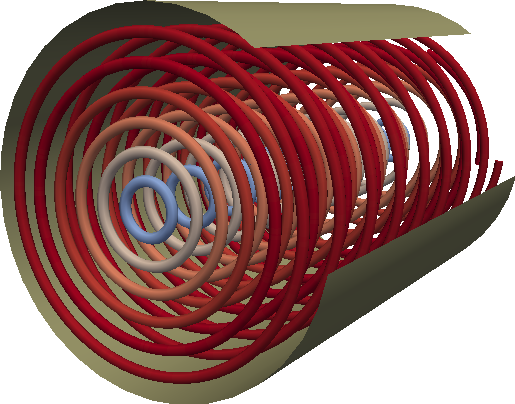}
		\caption{Magnetic field lines.}
	\end{subfigure}
	\caption{Field lines of the fundamental TM mode in a pillbox cavity.}
	\label{fig:pillbox_fields}
\end{figure}

As a numerical example with an available closed-form reference solution, we consider a pillbox (i.e. cylindrical) cavity with uncertain radius. We consider a fixed length $l=\SI{10}{\centi\meter}$ and we assume a uniformly distributed radius \mbox{$r\sim\mathcal{U}(\SI{4}{\centi\meter}, \SI{6}{\centi\meter})$.} Second degree basis functions are chosen for a total of \num{21692} degrees of freedom. This choice is a compromise between increased \mbox{regularity ($C^1$)} and sparseness and condition of the system matrix. The pillbox cavity as well as the electric and magnetic field lines of the fundamental \gls{tm} mode are shown in Fig.~\ref{fig:pillbox_fields}.

\subsubsection{Tracking}
We use the proposed eigenvalue tracking technique from section \ref{sec:tracking} to track the eigensolutions over the parameter range $r \in [\SI{4}{\centi\meter},\SI{6}{\centi\meter}]$.
The exact eigenfrequencies are well known \cite{Jackson_1998aa}. At a radius of approximately $r \approx \SI{4.92}{\centi\meter}$ a transition of the fundamental mode from \gls{te} to \gls{tm} is observed. 

\begin{figure}[t]
	\begin{subfigure}[c]{0.5\textwidth}
		\centering
		\includegraphics[]{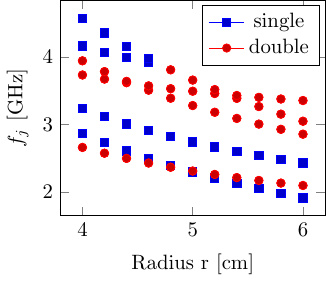}
		\caption{Separate computation of eigenvalues. Legend indicates multiplicity of modes.}
		\label{fig:pillbox_discrete_eigenvalues}
	\end{subfigure}
	\begin{subfigure}[c]{0.5\textwidth}
		\includegraphics[]{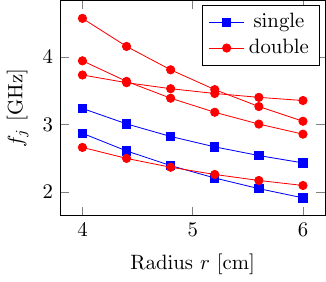}
		\caption{Tracking, markers show sample points. Legend indicates multiplicity of modes.}
		\label{fig:pillbox_traced_eigenvalues}
	\end{subfigure}
	\caption{Lowest 10 eigenfrequencies (including degenerate modes) of a Pillbox cavity with radius $r$ at discrete sample points (a) and using automatic tracking (b). The tracking follows only those 10 frequencies present at $r=\SI{6}{\centi\meter}$. This leads to the expected discrepancy in the computed frequencies $f_i$ e.g. at $r=\SI{4}{\centi\meter}$.}
\end{figure}

To illustrate the relevance of eigenvalue tracking, the eigenvalues obtained by solving \eqref{eq:galerkin-substitution-p} at discrete sample points $r=r_i$ are shown in Fig.~\ref{fig:pillbox_discrete_eigenvalues}. Without tracking, the eigenvalues cannot be associated to specific modes without cumbersome post-processing procedures that typically require some heuristics~\cite{Brackebusch_2014aa}.

For this first example, the algebraic homotopy \eqref{eq:homotopy} is replaced by the \textit{physical} system matrices $\ensuremath{\mathbf{K}} (r), \ensuremath{\mathbf{M}} (r)$ such that all eigenvalues computed along the homotopy path correspond to the solutions of a physical problem. This allows for the comparison with the closed-form solution with respect to the radius $r$. In practice, one is probably not interested in the intermediate points. However, in this test case the derivatives $\ensuremath{\mathbf{K}}', \ensuremath{\mathbf{M}}'$ cannot be straightforwardly computed and are then approximated by finite differences. Fig.~\ref{fig:pillbox_traced_eigenvalues} demonstrates the result of the tracked eigenvalues.

\textit{Remark: In section~\ref{sec:tracking} we assumed, for simplicity, that the eigenmodes at $t=0$ are not degenerate. Obviously, this assumption is not fulfilled for this test case. However, the proposed algorithm can still be applied since the degeneracy of the respective eigenvalues is caused by symmetry of the geometry which is preserved for all parameter changes.}

\subsubsection{Uncertainty Quantification}

We determine the uncertainty in the 6 lowest eigenfrequencies at $r=\SI{5}{\centi\meter}$ for the uniformly distributed radius $r\sim\mathcal{U}(\SI{4}{\centi\meter}, \SI{6}{\centi\meter})$ using stochastic collocation based on Clenshaw-Curtis quadrature with $N_\text{k} = 5$ collocation points. The eigenvalue problem for the first collocation point at $r=\SI{5}{\centi\meter}$ is solved based on Arnoldi's method \cite{Lehoucq_1996aa} using the command \textsc{eigs} in Matlab R2017a.  For each eigenpair the tracking method is applied for every remaining collocation point. The initial stepsize is chosen such that all points are reached using only one step. For this particular model, the Newton scheme always converges, using on average 2.2 iterations (max. 4) such that the stepsize is never reduced.

\begin{figure}[t]
	\centering
	\includegraphics[]{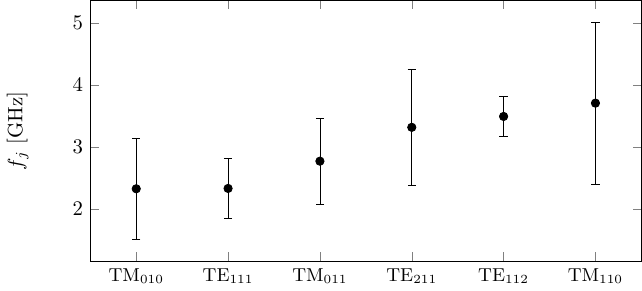}
	\caption{Expectation and $3\sigma$-interval for eigenfrequencies of pillbox cavity.}
	\label{fig:pillbox_uq}
\end{figure}

The first two statistical moments can be seen in Fig.~\ref{fig:pillbox_uq}. The relative errors in the numerical results with respect to the analytic reference solutions are below $3.5\cdot 10^{-4}$ (for expectations as well as standard deviations).

\subsubsection{Computational Cost}
The tracking is necessary to obtain correct results since the eigenmodes need to be matched at each collocation point. However, as shown in this subsection, it is also numerically more efficient than naively solving the repeated eigenvalue problems independently of each other. The computations are performed on a workstation with an Intel(R) Xeon(R) CPU E5-2687W \SI{3.1}{\giga\hertz} processor and 256 GB RAM. All time measurements are repeated $10$ times and averaged.  It shall be noted that, for a fair comparison, in all eigenvalue solvers the occurring linear systems of equations were solved with the Matlab \textit{backslash} operator \textbackslash, i.e. sparse Gaussian elimination. Improvements are to be expected when reusing factorizations or preconditioners with appropriate iterative solvers.

As a reference, we solve the eigenvalue problem at each collocation point using a Matlab implementation of the Jacobi-Davidson algorithm \cite{Sleijpen_1996aa,Fokkema_1998aa} as well as the Matlab command \textsc{eigs} which internally calls the \textsc{Arpack} library \cite{Lehoucq_1998aa}. Since we are interested in quantifying uncertainties for 10 eigenmodes and given that the ordering of eigenvalues may change, we calculate 20 eigenvalues to ensure that the corresponding eigenvalues are included at each collocation point.
The Jacobi-Davidson algorithm solves the eigenproblems on average in $42.0$ iterations and $\SI{931.4}{\second}$ while \textsc{eigs} takes on average \SI{192.6}{\second} for solving $80.6$ linear systems.

The computational cost of the proposed tracking approach is determined by the time needed for solving \eqref{eq:eigprobderiv} in order to compute the derivatives and for solving the linear systems of the Newton scheme \eqref{eq:NewtonLGS}. Cost for assembly of the system matrices is excluded since the algebraic mapping \eqref{eq:mat_derivatives} can be employed for \gls{uq}  which does not require any additional matrix assemblies. On average, the computational time needed for solving $3.2$ linear systems for each collocation point and eigenpair is \SI{16.0}{\second}.

\subsubsection{Shape morphing}
\begin{figure}[t]
	\begin{subfigure}[c]{0.29\textwidth}
	\includegraphics[height=0.12\textheight]{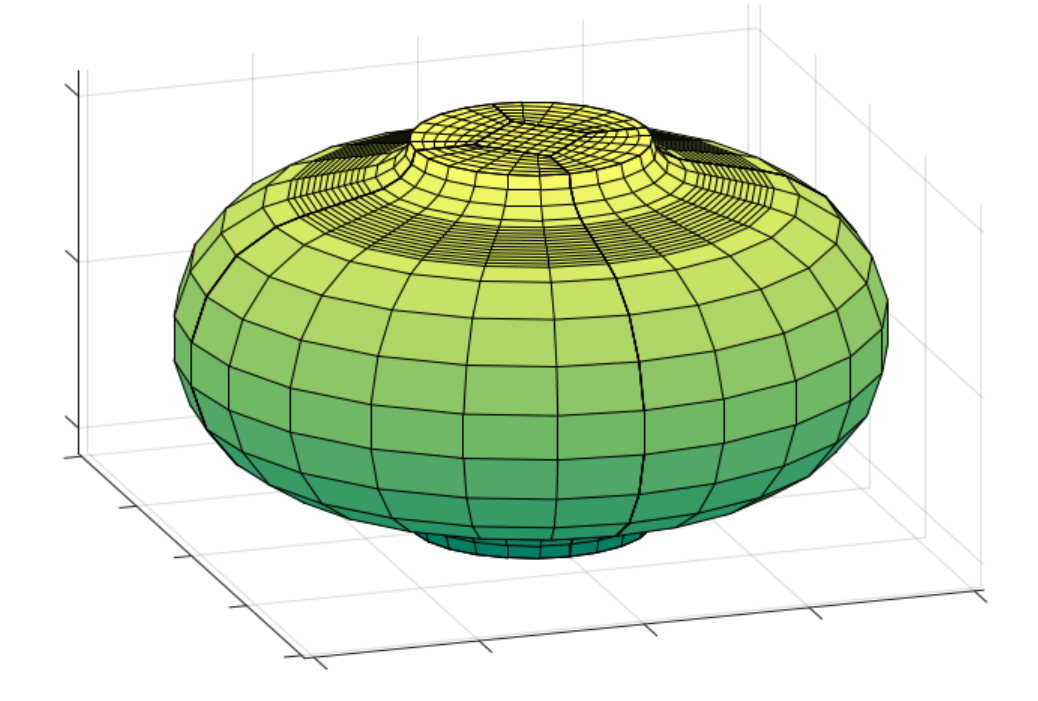}
	\caption{$t=0.0$}
	\end{subfigure}
	\begin{subfigure}[c]{0.29\textwidth}
		\centering
		{\includegraphics[height=0.12\textheight]{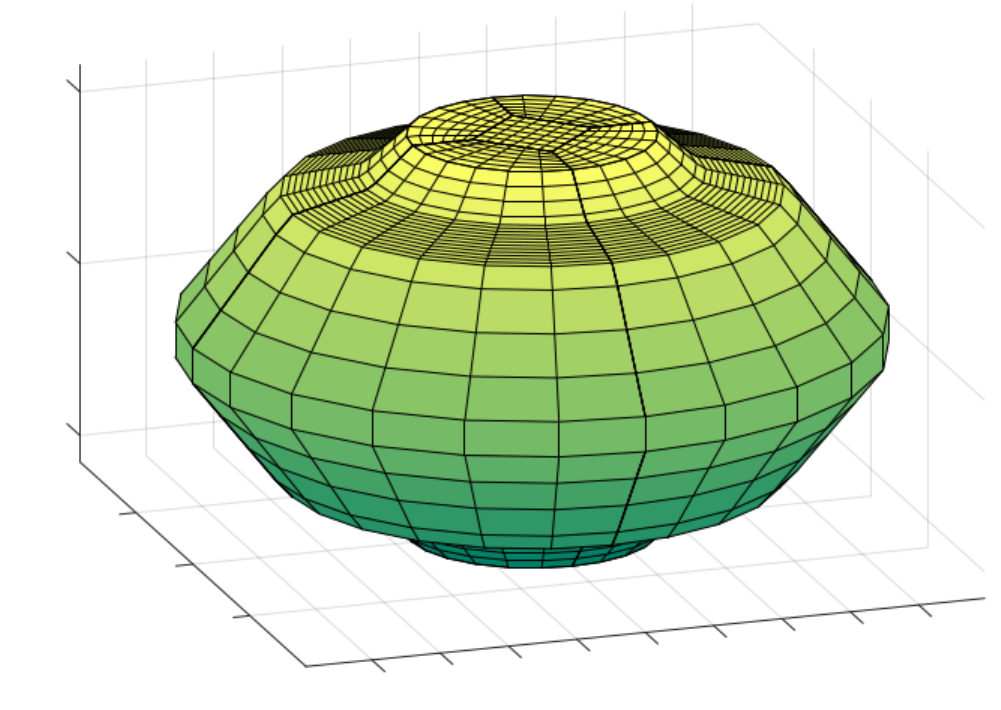}}	
		\caption{$t=0.5$}
	\end{subfigure}
	\begin{subfigure}[c]{0.22\textwidth}
		\centering
		{\includegraphics[height=0.12\textheight]{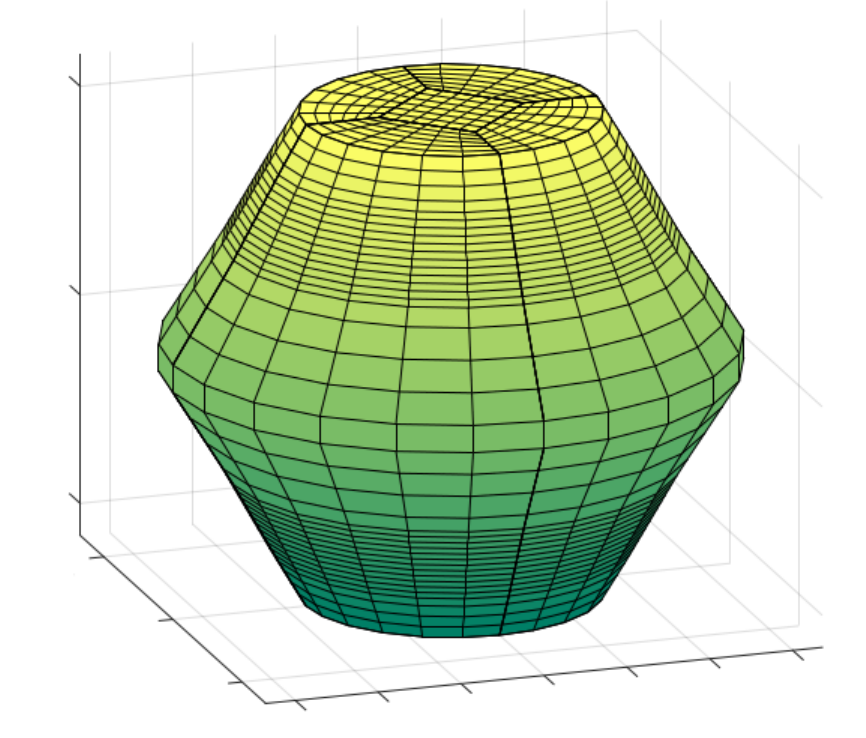}}	
		\caption{$t=0.75$}
	\end{subfigure}
	\begin{subfigure}[c]{0.17\textwidth}
		\centering
		{\includegraphics[height=0.12\textheight]{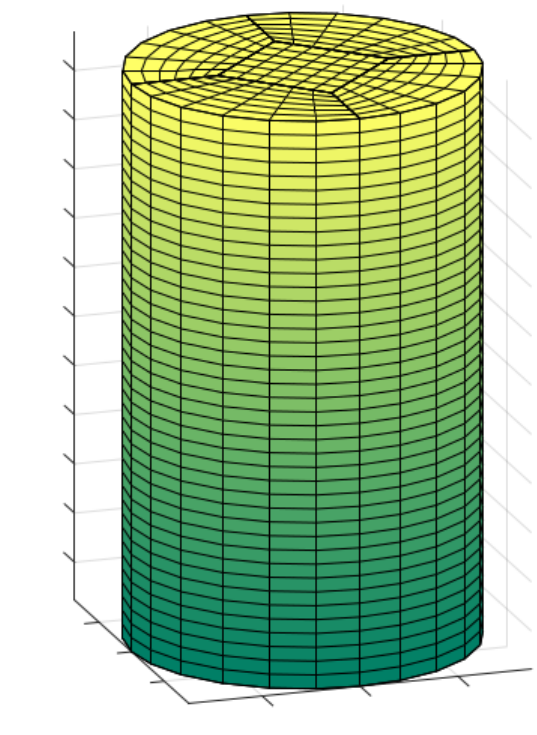}}	
		\caption{$t=1$}
	\end{subfigure}
\caption{Shape morphing of 1-cell TESLA cavity to pillbox cavity.}
\label{fig:morphing}
\end{figure}

\begin{figure}[h]
	\includegraphics[]{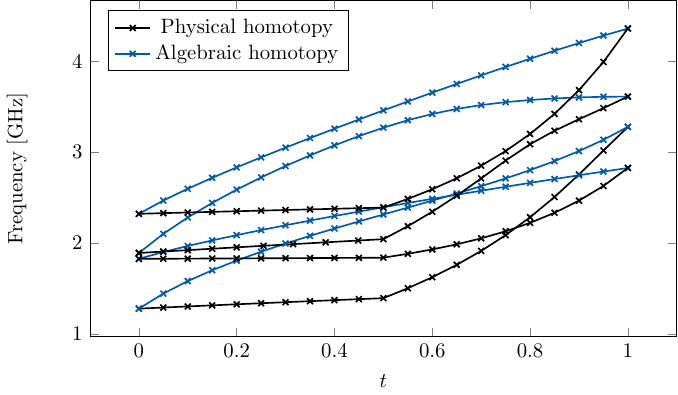}
	\caption{Tracking of eigenvalues during shape morphing of a \mbox{1-cell} TESLA  \mbox{cavity ($t=0$)} to a pillbox cavity ($t=1$).}
	\label{fig:MorphingTracking}
\end{figure}

In order to illustrate the ability of tracking the eigenmodes during more sophisticated deformations, we consider the shape morphing from the \mbox{1-cell} TESLA cavity to a pillbox cavity (radius $r=\SI{3.5}{\centi\meter}$, length $l=\SI{11.54}{\centi\meter}$), as illustrated in Fig. \ref{fig:morphing}. The eigenmodes in each cavity are discretized using second degree basis functions and \num{20712} degrees of freedom.
For this test case, we consider on the one hand the algebraic homotopy \eqref{eq:homotopy}  and on the other hand the \textit{physical} system matrices $\ensuremath{\mathbf{K}} (t), \ensuremath{\mathbf{M}} (t)$ corresponding to the depicted geometries $\forall t\in[0,1]$. In both cases, we apply the proposed tracking algorithm for the first few eigenvalues of the TESLA cavity. Fig.~\ref{fig:MorphingTracking} demonstrates the results of the tracked eigenvalues. It can be observed that both approaches match the same eigenmodes.

\subsection{TESLA Cavity}
A more challenging application is the simulation of a so-called \gls{tesla} type cavity. This is an \gls{rf} superconducting cavity for linear accelerators with a design accelerating gradient of $E_\text{acc}=\SI{25}{\mega\volt\per\metre}$. In particular, we focus on the \gls{ttf} design of the cavities \cite{TTF_1995} which were installed at \gls{desy} in Hamburg.

The TESLA cavity is a 9-cell standing wave structure of about \SI{1}{\metre} in length, whose accelerating mode resonates at \SI{1.3}{\giga\hertz}. The cavities are produced, one half-cell at a time, by deep drawing of pure niobium sheets \cite{TTF_1995}. The half cells are then connected together by \gls{ebm}. The production process inherently introduces geometry imperfections that need to be taken into account in operation. Here, we focus on the misalignment of each of the 9 cells with respect to the ideal axis of the cavity. The misalignment can produce undesired effects on the field quality and on the resonating frequency.

We consider \num{18} variables, corresponding to the displacement of the nine cell's centres in the $x$ and $y$ direction and we collect approximately \num{700} observations available from DESY measurements~\cite{Gall_2013aa} in the matrix $\mathbf{T}$ (see Eq. \ref{eq:obs_mat_T}). The deformations are normally distributed, but they also present cross correlations. To obtain uncorrelated variables for the \gls{uq} process, and to reduce the size of the parameter space, the truncated Karhunen–Loève decomposition introduced in section~\ref{sec:kl} is applied. This allows for the reduction from \num{18} to \num{7} variables according to the relative information criterion
\begin{align}
\frac{\sum_{i=1}^{N_\text{t}=7} \sigma_i}{\sum_{i=1}^{N=18} \sigma_i} \ge 0.95.
\end{align}

\begin{figure*}[t]
\includegraphics[]{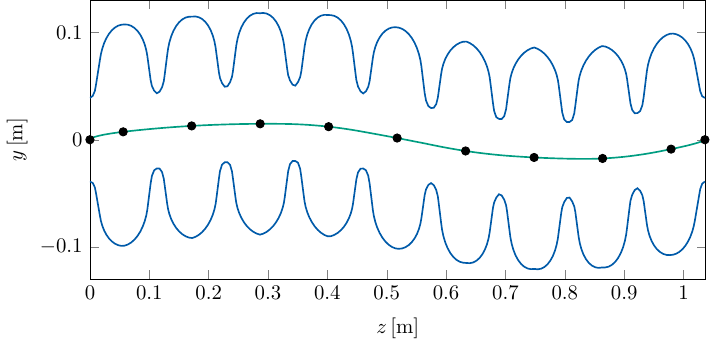}
\caption[Outline of a TESLA cavity with eccentric cells]{Outline of a TESLA cavity with eccentric cells (the two end beampipes are excluded from the picture for clarity). The measured deformation is enhanced by a factor of \num{500} for visualization reasons. The black points illustrate the offsetted centres of the cells, while the green line represents  the interpolating curve $\ensuremath{\mathbf{F}}_{\ensuremath{\boldsymbol\delta}}$. The final deformed cavity  $\ensuremath{\Omega\left(\ensuremath{\boldsymbol\delta}\right)}$ is given in blue.}\label{fig:tesla_eccentric_deformation}
\end{figure*}

The collocation points for the numerical quadrature are taken from a Smolyak grid of level \num{2}, with \num{3} Gaussian points along each direction, for a total of \num{127} points \cite{Nobile_2008aa}. For a given collocation point $\ensuremath{\boldsymbol\delta}_k$ it is possible to use Karhunen–Loève expansion and the curves $\ensuremath{\mathbf{F}}^{\mu}$ and $\ensuremath{\mathbf{F}}^{j}$ to build the deformation curve $\ensuremath{\mathbf{F}}_{\ensuremath{\boldsymbol\delta}_k}$ and thus the deformed geometry $\ensuremath{\mathbf{F}_{\ensuremath{\boldsymbol\delta}}}$. In Fig.~\ref{fig:tesla_eccentric_deformation} the deformed domain is shown.

\begin{figure}[t]
\centering
\includegraphics[]{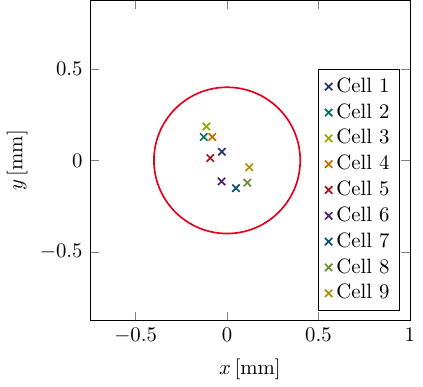}
\caption[TESLA cavity: cells eccentricity measurements]{TESLA cells eccentricity measurements of a particular cavity. Red circle illustrates the manufacturing tolerance.}\label{fig:tesla_eccentric_measurements}
\end{figure}

The measured eccentric deformations, e.g. shown in Fig.~\ref{fig:tesla_eccentric_measurements}, cause frequency shifts in the fundamental passband below $\SI{1}{\kilo\hertz}$. These shifts are rather harmless and can be corrected during the initial compulsory tuning procedure \cite{Padamsee_2008aa}. For the following numerical study, the magnitude of the deformations is artificially increased by a factor of 50 to demonstrate that our IGA workflow can cope with arbitrary deformations while linearization or ad-hoc mesh deformation will fail. The increased deformations lead to a frequency shift for the fundamental passband in the order of $\sim \SI{10}{\kilo\hertz}$, a level of accuracy that can be easily achieved by the isogeometric method by choosing second degree basis functions and approximately \num{600000} degrees of freedom.

\begin{table}
\center
\begin{tabular}{@{}lrrr@{}}
\toprule
Ref. Lev. & \ensuremath{N_{\text{dof}}} & $f_{\mathrm{TM_{010}}}$ [GHz] & $\Delta$ [kHz]\\
\midrule
\num{1} &  	\num{7772} & \num{1.313948189} &  -\\
\num{2} &  \num{24960} & \num{1.301478100} & \num{12470.089}\\
\num{4} & \num{111140} & \num{1.299939735} & \num{1538.365}\\
\num{6} & \num{299992} & \num{1.299971920} & \num{32.185}\\
\num{8} & \num{631836} & \num{1.299969453} & \num{2.467}\\
\bottomrule	
\end{tabular}
\caption{Computed frequency $f_{\mathrm{TM_{010}}}$ of the accelerating mode in the 9-cell TESLA cavity for increasing level of mesh refinement (IGA discretization with B-Spline basis functions of degree 2). The last column shows the incremental difference between subsequent refinement levels.\label{tab:refinement}}
\end{table}

\begin{figure}[t]
\includegraphics[]{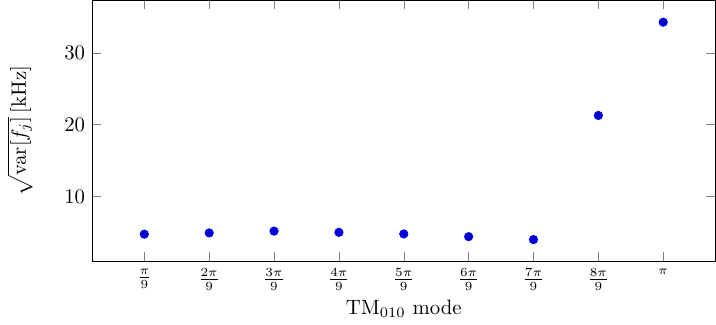}
\caption[TESLA cavity with eccentric cells: sensitivity of the TM$_{010}$ modes]{Standard deviation of the TM$_{010}$ mode in the TESLA cavity with eccentric cells. The results are obtained with deformations \num{50} times bigger than the measured ones.\label{fig:tesla_eccentric_stddev}}
\end{figure}

The results for the standard deviations of the \num{9} modes in the first TM$_{010}$ passband are depicted in Fig.~\ref{fig:tesla_eccentric_stddev}. It is noticeable, that the last two modes in the passband are more than \num{4} times more sensitive than the other ones. It shall be noted that, in practice, the frequency and the field distribution of the $9$-th TM$_{010}$ mode of all considered cavities are mechanically tuned on an automatic cavity tuning machine before installation. The impact of this process which should reduce the variability of the last modes in the first monopole passband has not been considered in this work and could be subject to future research.

\section{Conclusion}
In this work we investigate the application of \gls{uq} methods for the evaluation of eigenvalue sensitivity. In particular we are interested in the eigenfrequencies of \gls{rf} cavities since they are known to be highly sensitive to geometry deformations. 

Given the paramount role of geometry parameterization to achieve good accuracy, we discretize Maxwell's time-harmonic eigenvalue problem with \gls{iga}. The properties of the isogeometric basis functions also allow for a straightforward description of the domain deformation.

Finally, given that the ordering of the eigenfrequencies can change for different geometry configurations, we apply an eigenvalue tracking method, that exploits the properties of the \gls{iga} matrices, to ensure a correct classification of the modes in different collocation points. However, the tracking has the positive side effect of reducing the computational costs for solving the repeated eigenvalue problems.

\section*{Acknowledgment}
The work of J. Corno, N. Georg and S. Sch\"ops is supported by the \textit{Excellence Initiative} of the German Federal and State Governments and the Graduate School of Computational Engineering at Technische Universit\"at Darmstadt. J. Corno's and N. Georg's work are also partially funded by the DFG grants SCHO1562/3-1 and RO4937/1-1, respectively.

The authors would like to thank M. Kellermann and A. Krimm for their contributions to the first implementation and DESY for providing access to the cavity database.

\end{document}